\documentclass[sigconf]{acmart}
\AtBeginDocument{%
  \providecommand\BibTeX{{%
    \normalfont B\kern-0.5em{\scshape i\kern-0.25em b}\kern-0.8em\TeX}}}

\copyrightyear{2023}
\acmYear{2023}
\setcopyright{acmlicensed}\acmConference[CIKM '23]{Proceedings of the 32nd
ACM International Conference on Information and Knowledge
Management}{October 21--25, 2023}{Birmingham, United Kingdom}
\acmBooktitle{Proceedings of the 32nd ACM International Conference on
Information and Knowledge Management (CIKM '23), October 21--25, 2023,
Birmingham, United Kingdom}
\acmPrice{15.00}
\acmDOI{10.1145/3583780.3615483}
\acmISBN{979-8-4007-0124-5/23/10}

\newcommand{\ie}{\emph{i.e., }}
\newcommand{\eg}{\emph{e.g., }}

\newcommand{\etc}{\emph{etc.}}

\newcommand{\aka}

\usepackage{multirow} 
\usepackage{enumitem}
\usepackage{verbatim}
\usepackage[ruled]{algorithm2e}
\usepackage{multirow}
\usepackage{subfigure} 
\usepackage{ragged2e}
\usepackage{caption}
\usepackage{graphicx}
\usepackage[normalem]{ulem}
\useunder{\uline}{\ul}{}
\usepackage{amsmath}

\usepackage{bm}
\usepackage{multirow}
\usepackage{longtable}
\usepackage{adjustbox}
\usepackage{microtype}
\usepackage{setspace}
\usepackage{hyperref}
\usepackage{float}

\begin{document}
\newcommand{\bym}[1]{{#1}}
\newcommand{\zy}[1]{{#1}}

\title{Leveraging Watch-time Feedback for Short-Video Recommendations: A Causal Labeling Framework
}

\author{Yang Zhang*}
\orcid{0000-0002-7863-5183}
\affiliation{%
  \institution{University of Science and Technology of China}
  \city{Hefei}
  \country{China}
}
\email{zy2015@mail.ustc.edu.cn}
\thanks{*Equal contribution. Work done at Kuaishou.}

\author{Yimeng Bai*}
\orcid{0009-0008-8874-9409}
\affiliation{%
  \institution{University of Science and Technology of China}
  \city{Hefei}
  \country{China}
}
\email{baiyimeng@mail.ustc.edu.cn}

\author{Jianxin Chang*}
\orcid{0000-0002-7886-9238}
\affiliation{
 \institution{Kuaishou Technology}
 \city{Beijing}
 \country{China}}
\email{changjianxin@kuaishou.com}

\author{Xiaoxue Zang}
\orcid{0000-0002-5923-3429}
\affiliation{
  \institution{Kuaishou Technology}
  \city{Beijing}
  \country{China}}
\email{zangxiaoxue@kuaishou.com}

\author{Song Lu}
\orcid{0009-0009-0848-1391}
\affiliation{%
 \institution{Kuaishou Technology}
 \city{Beijing}
 \country{China}}
\email{lusong@kuaishou.com}

\author{Jing Lu}
\orcid{0009-0000-0718-6766}
\affiliation{%
 \institution{Kuaishou Technology}
 \city{Beijing}
 \country{China}}
\email{lvjing06@kuaishou.com}

\author{Fuli Feng$^{\dag}$}
\orcid{0000-0002-5828-9842}
\affiliation{%
  \institution{University of Science and Technology of China \& USTC Beijing Research Institute}
  \city{Hefei}
  \country{China}
}
\email{fulifeng93@gmail.com}
\thanks{$^{\dag}$ Corresponding author.}

\author{Yanan Niu}
\orcid{0000-0003-2083-518X}
\affiliation{%
 \institution{Kuaishou Technology}
 \city{Beijing}
 \country{China}}
\email{niuyanan@kuaishou.com}

\author{Yang Song}
\orcid{0000-0002-1714-5527}
\affiliation{%
 \institution{Kuaishou Technology}
 \city{Beijing}
 \country{China}}
\email{yangsong@kuaishou.com}

\def\authors{Yang Zhang, Yimeng Bai, Jianxin Chang, Xiaoxue Zang, Song Lu, Jing Lu,  Fuli Feng, Yanan Niu, Yang Song }

\renewcommand{\shortauthors}{Yang Zhang, et al.}

\begin{abstract}
  With the proliferation of short video applications, the significance of short video recommendations has vastly increased. Unlike other recommendation scenarios,  
  short video recommendation systems heavily rely
  on feedback from watch time. Existing approaches simply treat watch time as a direct label, failing to effectively harness its extensive semantics and introduce bias, thereby limiting the potential for modeling user interests based on watch time. To overcome this challenge, we propose a framework named \bym{Debiased} Multiple-semantics-extracting Labeling (DML). DML constructs labels that encompass various semantics by utilizing quantiles derived from the distribution of watch time, prioritizing relative order rather than absolute label values. This approach facilitates easier model learning while aligning with the ranking objective of recommendations.
  Furthermore, 
  we introduce a method inspired by {causal adjustment} to refine label definitions, 
  \bym{thereby directly mitigating bias at the label level.}
  We substantiate the effectiveness of our DML framework through both online and offline experiments. Extensive results demonstrate that 
  {our DML could effectively leverage watch time to discover users' real interests, 
  enhancing their engagement in our application.
  } 

\end{abstract}




\maketitle

\vspace{-5pt}
\section{Introduction}

Short-video content-sharing applications, such as Kuaishou and TikTok, have gained immense popularity worldwide~\cite{wang2019causes, directUse}, with their short-video recommendation systems playing a pivotal role in personalized video filtering~\cite{li2022billion, tang2017popularity}. These platforms provide various interfaces to elicit user preferences, including scrolling and commenting, which yield both explicit (e.g., likes) and implicit (e.g., watch time) feedback~\cite{wu2018beyond, D2Q}. However, due to the limited availability of explicit feedback, it is inadequate for building effective recommendation systems. Consequently, recommender system construction highly relies on dominant implicit feedback, particularly watch-time data. 
Hence, leveraging watch-time feedback to explore user interests becomes crucial in  enhancing user engagement and extending application usage. 

The existing research has primarily focused on using watch time as a direct label for data samples and improving the accuracy of watch time prediction. However, we argue that this approach of direct labeling fails to fully exploit the potential of watch time in modeling user interests. Unlike other implicit feedback signals like clicks, watch time contains valuable semantic information. For example, completing a video playback indicates a relatively strong user preference, which is not adequately emphasized by direct labeling methods. 
Additionally, using watch time as a direct label is susceptible to various biases, particularly duration bias, {as it is directly affected by factors beyond user-item matching (\eg video duration) without reflecting user preference~\cite{DCR,D2Q}. This necessitates debiasing techniques during training~\cite{WTG, D2Q}.}
However, current debiasing techniques often require modifying the model architecture or training methods~\cite{chen2023bias}, limiting their applicability to readily available recommender systems.

This study proposes a novel approach that utilizes watch time to generate explicit labels that capture different semantic aspects, enabling more accurate modeling of user preferences. The approach involves two types of labels: Watch-time Percentile Rank (WPR) labels and binary playback progress-related labels. The WPR label represents watch time but is discretized based on the percentile rank in the watch-time distribution, following the approach outlined by Roscoe~\cite{roscoe1975fundamental}. This discretization ensures label balance and robustness against outliers.
The binary playback progress-related labels are created by binarizing the playback progress using a threshold based on quantiles. By adjusting the quantile threshold, we can create labels that capture different semantics, such as whether a video has been effectively watched or completely watched, among others. Importantly, most of these labels are defined from a distribution perspective and emphasize relative ranking orders. 
\zy{This alignment with recommendation objectives facilitates effective learning.}

{However, these labels can potentially inherit the bias of watch time, derived from the impact of factors beyond user-item matching, such as the duration of the video. 
To address the bias issue, we propose a causal adjustment-inspired method~\cite{stratification} to refine the labels and mitigate the impact of these factors.} 
The method involves grouping videos based on biased factors and independently applying our labeling process to each group. Within each group, we assume that the influence of biased factors is negligible, as biased factors have an equal impact on all videos within the group. Additionally, our distribution-level definition of labels ensures that similar label distributions are generated for each group, regardless of the presence of biased factors. This allows us to disregard the impact of biased factors across groups. By mitigating bias during the labeling process, we eliminate the need for debiasing during model training, making these methods applicable to existing models. Considering the inclusion of debiasing in our labeling framework, we refer to it as \textit{\underline{D}ebiased \underline{M}ulti-semantics-extracting \underline{L}abeling} (DML).

The main contributions of this work are summarized as follows:
\begin{itemize} [leftmargin=*]
    \vspace{-10pt}
    \item 
    We propose the incorporation of multiple label categories reflecting the distribution of watch time, aimed at augmenting the utility of watch-time data. In addition, we introduce a {causality-based} methodology that refines these categories while simultaneously addressing potential bias.

    \item  
    We introduce a novel framework for relabeling and debiasing that transforms unprocessed organic user feedback into a range of training labels while directly countering label bias issues. This framework aims to maximize the exploitation of information-dense feedback generated by users.
     
    \item 
    \bym{We conduct extensive offline and online evaluations, effectively demonstrating the superiority of our method over baselines.}
\end{itemize} 

\section{Methodology}
In this section, we first provide an overview of our Debiased Multi-semantic-extracting Labeling framework, and then elaborate on the proposed labeling and debiasing method.

\begin{figure}[t]
  \centering
  \includegraphics[width=0.94\linewidth,height=4cm]{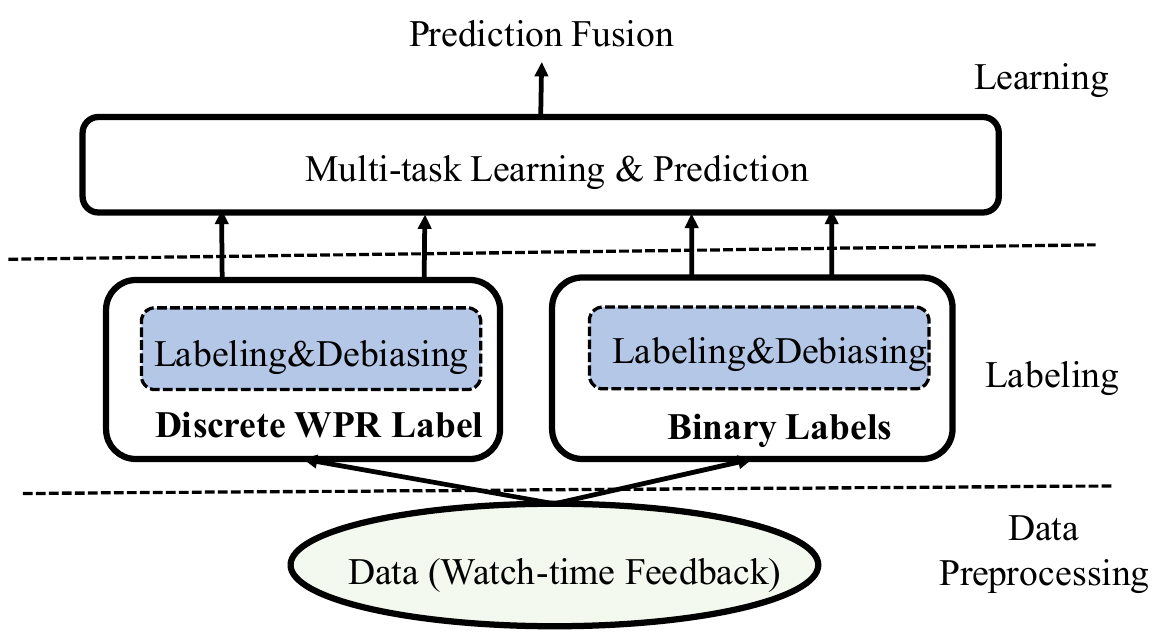}
  \vspace{-10pt}
  \caption{
  Overview of the system workflow for applying our DML framework, involving three main components: data pre-processing, labeling, and multi-task learning.
  }
  \label{fig:overall-framework}
  \vspace{-15pt}
\end{figure}

\subsection{Workflow Overview}
Within the context of short-video applications, the primary form of user feedback is the watch time for each video. This feedback is integral to the development of recommender models. Instead of using this feedback in its original form as a data label, we transform it into a spectrum of labels that underscore its multifaceted semantic dimensions. These labels are subsequently integrated into a multi-task learning architecture to precisely map user preferences. Furthermore, we directly implement debiasing during the labeling phase, thus eliminating the requirement for further debiasing during model training. This refinement enhances the practicability of our approach. The comprehensive system workflow, as illustrated in Figure~\ref{fig:overall-framework}, is divided into three principal components.

\vspace{+5pt}
\noindent $\bullet$ \textbf{Data pre-processing.}
\bym{This section is dedicated to discussing the user log pre-processing phase. Our focus is on refining data quality by reducing noise and applying feature engineering, including statistical feature generation. These initial steps are crucial for data preparation before advanced analyses or model generation.}

\noindent $\bullet$ \textbf{Labeling.}
This section emphasizes the transformation of watch time into discrete watch-time labels and binary labels related to playback progression, underscoring unique semantics. The discrete watch-time labels, representing watch time, are discretized and balanced and and can be utilized as labels for regression tasks. Contrarily, the binary labels related to playback progression indicate whether the user's video consumption has surpassed a specific threshold. Importantly, we incorporate debiasing directly into label generation at the label level, leveraging a causal methodology. This approach eliminates the necessity for debiasing during training, which often involves modifying model or training architectures.

\noindent $\bullet$ \textbf{Learning.}
Once we have obtained the debiased labels, we employ the Multi-gate Mixture-of-Experts (MMoE) architecture, a widely used architecture in recommendation, to construct our recommender model~\cite{ma2018modeling}. Each label serves as \bym{an} individual learning task, and we optimize the model to fit each task independently. By leveraging the diverse information captured by the different labels, we enhance the modeling of user interests.
During the serving phase, we generate multiple ranking lists using the predictions from individual tasks and the fused predictions. These diverse ranking lists serve as alternative recommendation routes.

The data pre-processing and learning components resemble existing work. Hence, the core novelty lies in our labeling component. We now proceed to describe the methodology for acquiring the two distinct types of debiased labels within our labeling framework.

\vspace{-5pt}
\subsection{Debiased Multi-Semantic-Extracting Labeling Framework}
In this section, we explain the proposed discrete watch time label  (Watch-time Percentile Rank) and provide details on binary labels.

\subsubsection{Watch-time Percentile Rank (WPR) } \label{sec:WPR}
Watch time has the potential to serve as a natural indicator. However, it often exhibits a substantial range and follows a characteristic long-tail distribution. In our scenario, the majority of videos exhibit a watch time of less than 120 seconds. However, there exist certain outliers that might extend beyond an hour, attributable to factors such as repetitive playback or association with particular video genres. To tackle this challenge, we propose the transformation of watch time into a label called ``\textit{Watch-time Percentile Rank}" (WPR). 
This new label provides a balanced and discrete representation, effectively mitigating the impact of outliers.

\vspace{+5pt}
\noindent $\bullet$ \textit{\textbf{Definition of WPR}}.
The concept of utilizing quantiles as a robust measure against outliers, as demonstrated by Aravkin et al.~\cite{aravkin2014orthogonal}, serves as an inspiration for our approach in transforming watch time. We propose the use of special quantiles known as percentiles to convert watch time into a discrete and range-limited label called "Watch-time Percentile Rank" (WPR). The WPR is defined as the percentile rank~\cite{roscoe1975fundamental} of a specific watch time within the overall population. It signifies the percentage of all watch times that are equal to or lower than the given watch time in its frequency distribution. As a result, the WPR label does not solely focus on the absolute value of watch time, but rather on its relative rank within the entire population.
To generate labels for training examples, we follow these  four steps: 

\begin{itemize}[leftmargin=*]
    \item[---] \textbf{Step 1}:  We rank all training examples based on their watch time feedback in ascending order.
    \item[---] \textbf{Step 2}: We choose the number of groups $N$, and set the ratio of each group relative to the entire dataset as ${q_{n}}$ ($n\in [1, N]$), subject to the constraint that  $q_{1}+ \dots + q_{N}=1$. 
    \item[---] \textbf{Step 3}: We split the ranked data into  $N$ groups  with the order kept, such that the ratio of data falling in $n$-th groups is equal to $q_{n}$, \ie  $\frac{|\mathcal{D}_{n}|}{|\mathcal{D}|}$ = $q_{n}$. Here $\mathcal{D}_{n}$ denotes the data that falls in the $n$-th group and $\mathcal{D}$ denotes all data.  
    \item[---] \textbf{Step 4}: For each training example $(x,y)$ in $D_{n}$, we take $y_{wpr} = \sum_{n^{\prime}=1}^{n} q_{n^{\prime}}$ as the WPR label\footnote{Original percentile rank would multiply by 100, \ie $y_{wpr}\times100$. Here we omit "$\times 100$".}.
\end{itemize}
Let us define the label generation process using the function $g_{wpr}(\cdot)$. For each $(x,y)\in \mathcal{D}$, the transformed label is computed as 
\begin{equation}\label{eq:wpr-biased}
    y_{wpr}=g_{wpr}(x,y;\mathcal{D};\{q_{1},\dots,q_{N}\}),
\end{equation}
% $$$$
where $y$ denotes the original watch time. Notably, $y_{wpr}$ is constrained within the range of $[0,1]$ and can only take on one of $N$ possible prefix sum values, determined by the set $\{q_{1},\dots,q_{N}\}$.

In order to achieve a balanced distribution of labels, one approach is to set $q_{1}=q_{2}=\dots=q_{N}=1/N$. This configuration is referred to as the "native WPR" label. However, using this naive method can lead to densely populated labels in areas associated with short watch times. For example, in the Kuaishou scenario with $N=300$, there are 41 unique label values within the range of 0-3 seconds, which is the same as the range of 40-90 seconds. Treating each possible label equally during training can result in the model exerting the same effort to differentiate between watch time differences of 0-3 seconds and 40-90 seconds, which is not desirable.

\zy{To address this issue, we adopt a progressive finer-grained partitioning strategy, where the intervals between labels within short watch time are set larger, \ie we enforce $q_{1}\geq q_{2}\geq \dots \geq q_{N}$. } The specific values of $q_{n}$ are determined based on empirical experience\footnote{In our scenarios, we select $\{q_{1},\dots,q_{N}\}$, making the watch time ($y$) and the WPR ($y_{wpr}$) satisfy: 
$1/y_{wpr} \approx a \lceil ln(y) \rceil^2 + b \lceil ln(y) \rceil + c$, where $a,b$ and $c$ are hyper-parameters and determined through empirical experience.}.

\begin{figure}[t]
\centering
\includegraphics[width=0.35\textwidth]{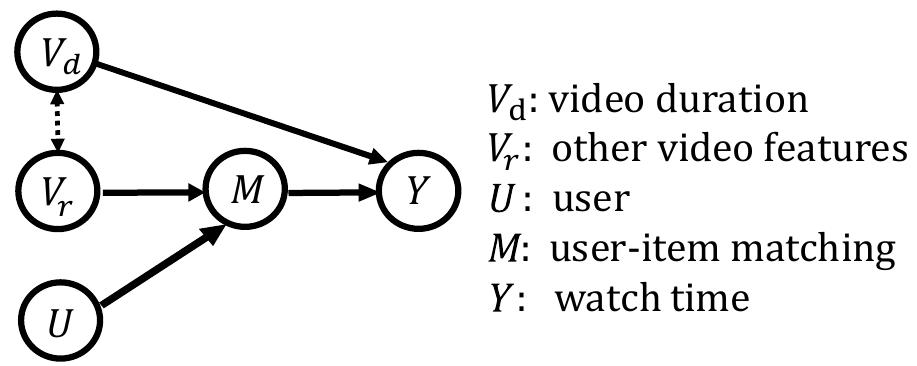}
\vspace{-5.5mm}
\caption{{
Causal graph to describe the generation process of watch time feedback. 
$Y$ is not only affected by $M$, but also directly by $V$ without reflecting user preference. 
}}
\vspace{-15pt}
\label{fig:causal_graph}
\end{figure}

\noindent $\bullet$ \textit{\textbf{Label Debiasing}}. 
As discussed by previous work~\cite{DCR,D2Q}, the watch-time feedback is not only affected by user-item matching  but also directly by the video duration. We could abstract the causal relations with the causal graph in Figure~\ref{fig:causal_graph}, with $M \rightarrow Y$ and $V_d \rightarrow Y$ to represent the influence of user-item matching $M$ and the direct influence of video duration $V_d$ on $Y$. Here we assume the matching of a video to $U$ is determined by video features $V_r$~\cite{DCR,D2Q},
representing by $(U, V_r)\rightarrow M$; and we take $V_{d}\dashleftrightarrow V_{r}$ to represent the item features could potentially affect each other or have a common cause, \eg the video creator~\cite{DCR}. According to the graph, $V_{d}$ is a confounder between $V_{r}$ and $Y$, its confounding effects would result in bias. Meanwhile, its direct effects are entangled with the effects of user preference, which would mislead the learning of user preference, bringing bias. If directly creating the WPR, it undoubtedly inherits such biases.   

To mitigate such biases, the key lies in cutting down the edge $V_{d}\rightarrow Y$ to model the causal effect of M, which could be estimated with the causal adjustment
~\cite{DCR,D2Q} as follows: 
\begin{equation}\small
    \sum_{v_d \in \mathcal{V}_d } P(Y|U,V_r,v_d)P(v_d),
\end{equation} 
where $\mathcal{V}_d$ denotes the all possible video duration. 
The adjustment works with two mechanisms~\cite{DCR,pearl2009causality}.
Firstly, for sub-populations with the same video duration $V_d = v_d$, the influence of $v_d$ is assumed to be the same. Within each sub-population, relative user preference can be directly obtained through prediction result comparison. Secondly, all possible values of $V_d$ are enumerated with the same $P(V_d)$ for each sub-population in a counterfactual manner, which ensures that the influence of $v_d$ is the same across all sub-populations. This enables the relative user preference could also be obtained via direct comparison across sub-populations.

We try to simulate the adjustment process during generating the WPR label to achieve debiasing. To achieve this, we first split the interactions into different groups according to the video duration. Then, we simulate the first mechanism of the adjustment by independently running the WPR generation process for each group. To simulate the second mechanism of the adjustment, it is hard to forcibly change the video duration during generating the label to perform enumeration. However, for each video duration group, the generated WPR label obeys the same distribution. We thus could roughly think that the influence of the video duration is the same over different groups, achieving the similar result of the second step of the adjustment. 
Formally, let $\mathcal{D}^{v_d}$ denote the group data with the video duration of $v_d$, for each training example $(x,y)$ in $\mathcal{D}^{v_d}$, we generate the debiased WPR label as follows:
\begin{equation}
    y_{wpr}^{d} = g_{wpr}\left(x,y;\mathcal{D}^{v_d};\{q_{1},\dots,q_{N}\}\right),
\end{equation}
where $x$ could also be represented by $(u,v,v_d)$. Compared to the original generation process, the only difference is that we perform the process in $\mathcal{D}^{v_d}$ instead of $\mathcal{D}$.

\subsubsection{Binary Labels}
We next consider creating binary labels to emphasize whether the watching of videos reaches a specific degree, such as whether a video has been effectively watched. To create such labels, we binarize the watch time with a specific percentile threshold. 
By varying the threshold, we could create different labels with different semantics. Here, we mainly introduce two labels: 

\vspace{+5pt}
\noindent \textbf{Effective View (EV).} This label is used to reflect whether a video has been effectively viewed. Specifically, if the watch time on a video exceeds that of 50\% of training examples, we think the video is effectively viewed. Formally, for a training example $(x,y) \in \mathcal{D}$, we define the effective view label $y_{ev}$ as follows:
\begin{equation}
    y_{ev} = \mathbb{I}(y \geq t_{50}(\mathcal{D})), 
\end{equation}
where $\mathbb{I}(\cdot)$ denotes the indicator function, and $t_{50}(\mathcal{D})$ denotes the 50th percentile (\ie median) of watch time about $\mathcal{D}$.

\vspace{+5pt}
\noindent \textbf{Long View (LV).} Similarly, we define the long view label to reflect whether the watch time of a training example has exceeded 75\% of watch time in the training dataset. Formally, for a training example $(x,y)$ in $\mathcal{D}$, the long view label $y_{lv}$ is defined as:
\begin{equation}
    y_{lv} =  \mathbb{I}(y \geq t_{75}(\mathcal{D})), 
\end{equation}
where $t_{75}(\mathcal{D})$ denotes the 75th percentile of watch time for $\mathcal{D}$.

\vspace{+5pt}
\noindent $\bullet$ \textit{\textbf{Label Debiasing.}} It is evident that the EV and LV labels also suffer from  duration bias, such as {longer videos watched by users are easier to obtain positive labels. We next present how to mitigate the bias for these labels}. 

\noindent \textbf{Duration Bias.} As the EV and LV are generated based on watch time, the source of duration bias should be the same as that for the WPR label. We directly apply the same debiasing strategy, \ie independently running the labeling process for groups split by the video duration. Taking the EV label as an example, for a training example $(x,y)$, we generate the debiased label as follows:
\begin{equation}
    y_{ev}^d = \mathbb{I}\left(y \geq t_{50}(\mathcal{D}^{v_d})\right), 
\end{equation}
where $v_d$ denotes the video duration of the sample,  $\mathcal{D}^{v_d}$ denotes all data with video duration of $v_d$ in $\mathcal{D}$, and $t_{50}(\mathcal{D}^{v_d})$ denotes the median of watch time in $\mathcal{D}^{v_d}$.

\noindent \textbf{Other Biases.} According to previous work~\cite{zhang2021causal, wang2021deconfounded, wei2021model}, other user-side or item-side factors like video popularity could also play a similar causal role to the video duration, bringing bias. For the WPR label, we believe that duration bias plays the dominant role, and we thus ignore the biases brought by these biased factors. 
However, the EV and LV labels could be more susceptive to biases. 
For example, the EV and LV label of some videos/users could easily always be negative\footnote{We usually do not suffer from the issue for the WPR label, since we take a more fine-grained labeling strategy for it.} even after mitigating duration bias, especially for unpopular videos/inactive users. This would result in over-suppressing unpopular videos and unsatisfying inactive users.

To address this issue, a similar debiasing strategy to duration bias can be adopted, but a more fine-grained group split is needed based on the combination of all possible biased factors~\cite{pearl2009causality}. However, such biased factors may be difficult to detect fully, and splitting based on their combination could result in very sparse groups. To overcome this challenge, we propose to use all available features of a user or video as a substitute for such factors and separately perform  debiasing on the user side and item (\ie video) side.
Formally, for a training example $(x,y)$, we generate debiased labels from two sides as follows (with the EV label as  an example):
\begin{equation}
    \begin{split}
        \text{item side}: y_{ev}^v = \mathbb{I}(y\geq t_{50}(\mathcal{D}^{v})),\\
        \text{user side}: y_{ev}^u = \mathbb{I}(y\geq t_{50}(\mathcal{D}^{u})),
    \end{split}
\end{equation}
where $v$ denotes video features containing $v_d$ and $v_r$, \ie $v=[v_d,v_r]$, $\mathcal{D}^v$ ($\mathcal{D}^u$) denotes all samples with the video (user) features of $v$ ($u$), and $t_{50}(\mathcal{D}^v)$ ($t_{50}(\mathcal{D}^u)$) denotes its watch time median.
\section{Experiments}
In this section, we conduct both offline and online experiments on real industrial data to evaluate our proposed method, answering the following three research questions: \textbf{RQ1:} How is the performance of the proposed DML compared with existing methods? \textbf{RQ2:} How do the design choices in DML affect its efficacy? \textbf{RQ3:} Can the proposed DML improve real-time video consumption on the Kuaishou video recommendation platform?

\begin{table}[t]\small
\small
\caption{Statistics of the industrial dataset  constructed on the daily collected user logs from Kuaishou.}
\vspace{-10pt}
\resizebox{0.4\textwidth}{!}{%
\begin{tabular}{llc}
\toprule
Data & Field & Size \\ 
\midrule
\multirow{4}{*}{Daily Log Info} & Users & 345.5 million \\
& Videos & 45.1 million \\
& Samples & 46.2 billion \\
& Average User Actions & 133.7 / day \\
\midrule
\multirow{2}{*}{Historical Behaviors} & Average User Behaviors & 14.5 thousand \\
& Max User Behaviors & 100 thousand \\
\bottomrule
\label{tab:dataset}
\end{tabular}
}
\vspace{-15pt}
\end{table}

\vspace{-5pt}
\subsection{Experimental Settings} 

\noindent $\bullet$ \textit{\textbf{Data Information.}}
We conduct experiments  on industrial data from Kuaishou, one of the top short-video sharing platforms in China.
As shown in Table \ref{tab:dataset}, the size of daily active users on Kuaishou is around 346 million. 
Every day 45 million short videos are posted and these videos are played 46 billion times in total. On average, each user watches 133.7 short videos per day. 
To utilize rich behavior information, we collect full user historical behaviors from older logs back months ago. On average, each user watched 14,500 videos in the past six months. We cut off the maximum user behavior sequence length to 100,000, which is about the annual total number of views for heavy users.
Specifically, each data sample is associated with an important feature ``video duration'' denoting the total length of the video, and the feedback of ``watch time'' denoting the total seconds that the user played this video.

\noindent $\bullet$ \textit{\textbf{Baselines.}}
To verify the effectiveness of our DML, we compare it with the following methods that focus on leveraging watch time:
\begin{itemize}[leftmargin=*]
    \item[-] \textbf{TR~\cite{directUse2}.}
    This refers to the traditional regression method using watch time  as the label directly and learning a regression model by minimizing mean squared error (MSE).
    
    \item[-] \textbf{WLR~\cite{youtubeDNN}.} 
    This refers to Youtube's Weighted Logistic Regression, which learns a logistic regression model by fitting all interactions, reweighted by watch times, and uses the learned odds to estimate watch time during prediction.
    
    \item[-] \textbf{OR~\cite{ordinalR}.}  This refers to an approach that uses ordinal regression~\cite{ordinalR} to fit watch time, which places emphasis on the relative order of watch times during the learning process.
    
    \item[-] \textbf{D2Q~\cite{D2Q}.} 
    \zy{This is a SOTA method in watch time prediction. It uses backdoor adjustment to address duration bias and involves fitting duration-dependent quantiles of watch time using MSE.}
\end{itemize}
To ensure a fair comparison, we implement all methods within Kuaishou's multi-task learning framework, developed based on the MMoE algorithm~\cite{ma2018modeling} and including other tasks such as click prediction and complete watching prediction. 
For all methods, we employ the AdaGrad optimizer~\cite{lydia2019adagrad} in the embedding layer with a learning rate of 0.05. The DNN parameters are updated using the Adam optimizer~\cite{kingma2014adam} with a learning rate of 5e-6, and the batch size is set to 8192.  These choices are based on empirical observations. 

\begin{table}[t]\small
\caption{Offline results regarding watch time prediction, where the best results are highlighted in bold and sub-optimal results are underlined.}
\vspace{-10pt}
\label{tab:main}
\resizebox{0.45\textwidth}{!}{%
\begin{tabular}{cccccc}
\toprule
Methods           & AUC$\uparrow$      & GAUC$\uparrow$      & MAE$\downarrow$        & MAPE$\downarrow$     & RMSE$\downarrow$                           \\ \hline
TR & 0.6597 & 0.6397 & 24.1743 & 3.6892 & 45.4747
\\
WLR     & 0.6711 & 0.6551 & 23.5743 & 3.1528 & \uline{43.3776}                      \\
OR     & 0.6727 & 0.6474 & 22.8930 & 3.5017 & 44.3891                      \\
D2Q   & \uline{0.6732} & \uline{0.6581} & \uline{22.6728} & \uline{2.7342} & 45.5293                      \\ 
DML                                                  & \textbf{0.6763} & \textbf{0.6617} & \textbf{21.7657} & \textbf{2.6039} & \textbf{42.7735} \\ 
\bottomrule
\end{tabular}
}
\vspace{-5px}
\end{table}

\vspace{-5pt}
\subsection{Offline Evaluation (RQ1 \& RQ2)}
\textbf{Offline Setting}. 
We use samples collected over a 23-hour period as the training data, with samples collected during the subsequent hour as the testing data.
We evaluate watch-time prediction performance using two types of evaluation metrics: ranking metrics (AUC and GAUC, which is AUC averaged over users) and accuracy metrics (MAE, RMSE, and MAPE~\cite{MAPE}, which refers to Mean Absolute Percentage Error). We would convert the prediction results back into watch time when computing these metrics if quantile-based labels are employed, following~\cite{D2Q}.

\subsubsection{Performance Comparison (RQ1)}
The comparison results are summarized
in Table \ref{tab:main}, where we draw the following observations:
\begin{itemize}[leftmargin=*]
    \item 
    Our DML outperforms all baselines on all metrics, highlighting its effectiveness in unlocking the potential of watch time for modeling real user interest. This success can be attributed to its ability to extract diverse semantics contained in watch time and perform label debiasing.
    
    \item 
    D2Q outperforms other baselines in most of the metrics. This can be mainly attributed to its additional considerations for debiasing, which highlights the importance of addressing bias issues when leveraging watch time. However,  compared to DML, although D2Q also conducts debiasing, it consistently performs worse. This is because D2Q fails to consider the diverse semantics of watch time that DML could capture.

    \item TR exhibits the worst performance among all compared methods, likely due to its direct utilization of watch time as the sole label. Although OR utilizes watch time similarly, it outperforms TR, which may be attributed to its emphasis on the relative ranking order of watch time for learning user preference.
\end{itemize}

\begin{figure}[t]
    \centering
    \includegraphics[width=0.8\linewidth,height=2.7cm]{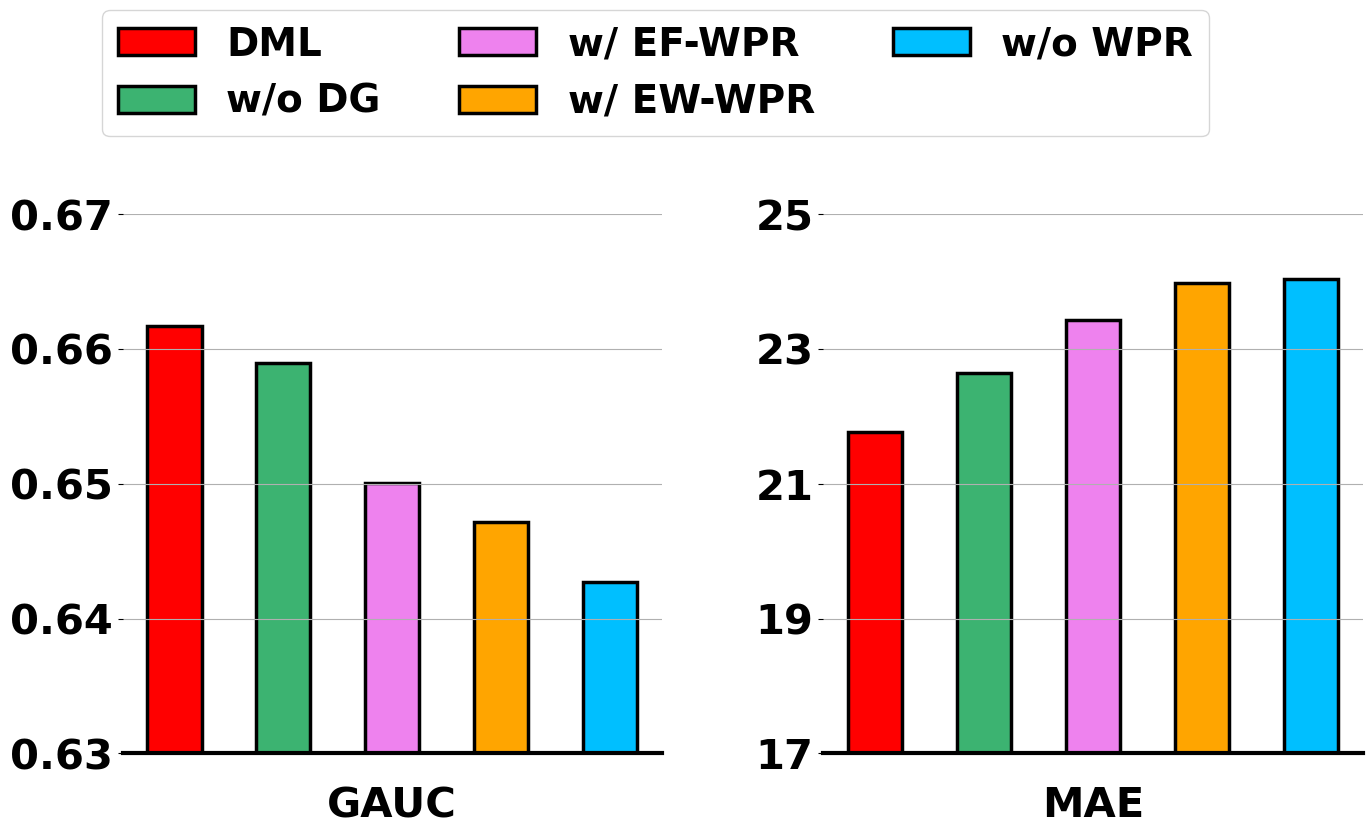}
    \vspace{-10pt}
    \caption{Ablation studies for the WPR label.}
    \label{fig:WPR}
    \vspace{-10pt}
\end{figure}

\begin{figure}[t]
    \centering
    \includegraphics[width=0.8\linewidth,height=2.7cm]{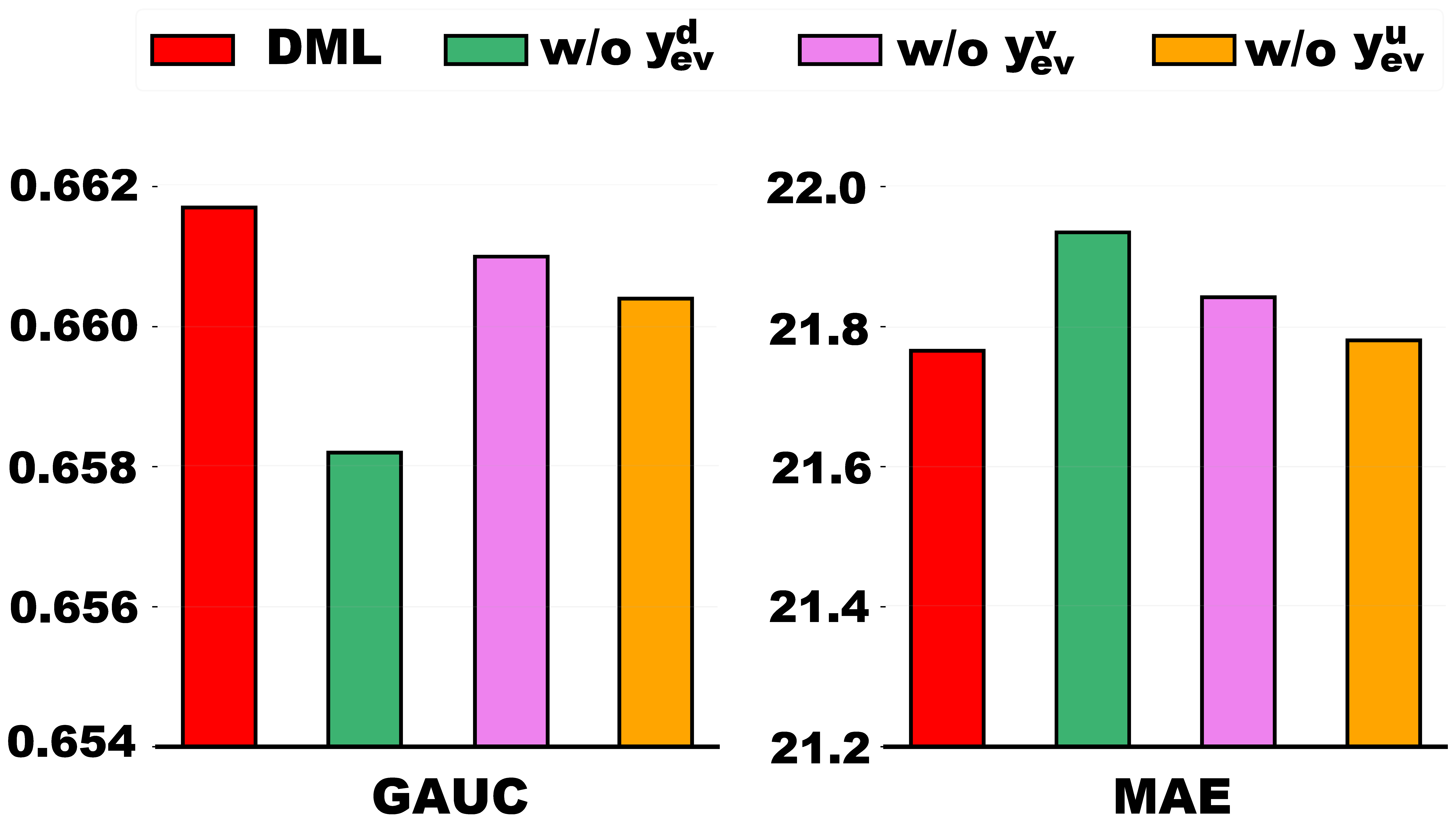}
\vspace{-10pt}
    \caption{Ablation studies for the binary labels.}
    \label{fig:binary}
    \vspace{-15pt}
\end{figure}

\subsubsection{Ablation Studies (RQ2)}
We next conduct experiments to study the influence of design choices in the WPR labels and the necessity of binary labels (taking the EV label as an example).

\vspace{+3pt}
\noindent \textbf{Studies on the WPR Label.}
There are two critical components when generating the WPR label: debiasing and progressive percentile splitting. 
Our debiasing relies on two main designs: 1) grouping based on duration and 2) using the percentile-based label to keep similar label distributions across different duration groups. 
To examine these designs, we conducted two variants of DML: DML without duration-based grouping (w/o DG), and DML with finishing playing rate (w/o WPR), which uses the playing rate label instead of the percentile-based label WPR. 
To verify the progressive percentile splitting, we introduce two additional variants: DML with equal-frequency WPR (w/ EF-WPR), \ie making $q_{k}=1/N$ in Equation~\eqref{eq:wpr-biased}, and DML with equal-width WPR (w/ EW-WPR), which makes each group of watch time span an equal width\footnote{The EW-WPR also cannot bring similar label distributions across duration groups.}. 
Figure \ref{fig:WPR} summarizes the results of DML and its variants on GAUC and MAE metrics. 

From the figure, we have two main observations. Firstly, we observe distinct declines in performance upon removing the duration-based grouping (w/o DG) or replacing the WPR with the playing rate (w/o WPR). These findings suggest that debiasing is crucial for capturing real user preference from watch time and that both debiasing designs are necessary. Secondly, the variants with equal-frequency WPR and equal-width WPR both perform worse than the original WPR. The equal-frequency WPR pursues a balanced label distribution but places excessive attention on shorter watch times, while DML w/ EW-WPR pursues equal treatment of different watch times without maintaining balance. The original WPR could strike a balance between the two variants, achieving better results.

\vspace{+3pt}
\noindent \textbf{Studies on the Binary Label.} 
We used the effective view (EV) label as an example to study the influence of binary labels. To better utilize the EV label, we primarily focus on resolving duration bias ($y_{ev}^{d}$), as well as other biases from the item side ($y_{ev}^{v}$) and user side ($y_{ev}^{u}$). 
To study the influence of these debiasing designs, we conduct experiments by removing the corresponding debiased labels, i.e., removing $y_{ev}^{d}$, $y_{ev}^{v}$, and $y_{ev}^{u}$, respectively. The results are presented in Figure \ref{fig:binary}.
It is evident that removing bias from each perspective is helpful for improving performance, with mitigating duration bias having the most significant influence on the effectiveness of DML.

\vspace{-5pt}
\subsection{Online Evaluation (RQ3)}
\textbf{Online Setting.}
We conduct an online A/B test to further evaluate the proposed method. Specifically, we run DML, WLR and D2Q, trained with offline data, for 5 consecutive days on Kuaishou's two recommendation channels: Featured-Video Tab and Slide Tab. We then report the average performance over the 5 days.
Here, we use WLR as the control group, and implement DML by gradually adding different labels to study their influence, where each implementation is denoted by "DML(labels)". For example, DML ($y_{wpr}^d$) means only using the debiased WPR label.
\zy{We utilize two evaluation metrics: Watch Time, measuring total user video consumption within a segment, and App Usage, offering insights into how frequently users access and engage with the Kuaishou App.}

\noindent \textbf{Performance Comparison.} 
\zy{The performance of the evaluated methods, as tested concurrently on the Kuaishou App, is presented in Table \ref{tab::abtest}, where we draw three observations.} 
Firstly, the improvement in performance upon utilizing the WPR label over WLR and D2Q demonstrates the effectiveness of utilizing our percentile rank-based labels. 
Secondly, the enhancement of both Watch Time and App Usage metrics when employing the debiased WPR label verifies the efficacy of the debiasing approach. 
Lastly, the incorporation of multiple EV or LV labels, addressing biases originating from video duration, user-side, and item-side factors, resulted in a rise in both metrics. \zy{This substantiates that these labels can capture distinct information and address various forms of biases.}

\vspace{-8pt}
\section{Related Work}

\noindent $\bullet$ \textbf{Watch-time Feedback Utilization.}
\bym{Watch-time feedback is a crucial signal for user preferences in video recommendation~\cite{WTG}.
Leveraging this feedback to build effective recommenders has gained significant attention~\cite{directUse,directUse2,directUse3, DCR, D2Q, youtubeDNN, WTG}.
Many of these approaches focus on creating a watch time prediction task to leverage this feedback ~\cite{youtubeDNN,directUse,directUse2,directUse3}.
For instance, the pioneering work WLR~\cite{youtubeDNN} use the feedback to adjust clicks' weights during training, implicitly predicting watch time. Others directly use the feedback as labels for training prediction models~\cite{directUse,directUse2,directUse3}, potentially introducing bias towards longer videos.
To address biases, DVR~\cite{WTG} introduces the Watch Time Gain metric, while D2Q~\cite{D2Q} uses quantile-based labels based on backdoor adjustment~\cite{pearl2009causality}.
Unlike D2Q, our method adopts a progressive label-splitting strategy to form percentile-based labels (Section~\ref{sec:WPR}).
%and doesn't require model changes.
Apart from watch time prediction tasks, some studies convert watch time into binary labels~\cite{finishplay, DCR, effective_view, binary_label}, focusing on complete video watching, overlooking other aspects.
In contrast, our approach captures diverse watch time semantics simultaneously, using multi-task learning to enhance recommendations mutually.
Among these works, only DCR~\cite{DCR} employs causal estimation to tackle biases but involves model modifications.
%, necessitating model adjustments. 
Unlike DCR, our method performs debiasing during label generation without modifying the model.}

\noindent $\bullet$ \textbf{Recommendation Debiasing.}
\bym{Recommender systems grapple with biases~\cite{chen2023bias}, including position bias~\cite{collins2018study, joachims2017accurately, joachims2007evaluating}, selection bias~\cite{hernandez2014probabilistic, marlin2012collaborative, steck2013evaluation}, and popularity bias~\cite{zhang2021causal,wei2021model} \etc~ Three key research avenues have emerged to address these biases.
The first approach centers on reweighting~\cite{schnabel2016recommendations, wang2019doubly, saito2020unbiased, seqdebias}, with inverse propensity scores~\cite{schnabel2016recommendations} as a representative technique. This method modifies the training distribution via reweighting to achieve debiasing. However, estimating weights can pose challenges~\cite{DCR}.
The second approach employs unbiased data for model learning~\cite{haoxuan-data,chen2021autodebias,unbiasData,bonner2018causal}, though obtaining such data is often costly.
A third avenue targets biases from a causal standpoint, categorized into intervention~\cite{wang2021deconfounded, yang2021top, zhang2021causal,SentimentDebias} and counterfactual methods~\cite{wei2021model, clickbait,ying2023camus}. The intervention method incorporates causal adjustments~\cite{zhang2021causal,D2Q} like backdoor adjustment or other deconfounding strategies~\cite{BeCausal,DCF}, such as representation balancing, to tackle bias from confounding~\cite{pearl2009causality}. The counterfactual method employs counterfactual reasoning to debias, typically by comparing factual and counterfactual scenarios~\cite{wei2021model,clickbait}. These methods generally perform debiasing during model learning, potentially altering model structures or training processes. In contrast, our causal method operates in full pre-processing, debiasing during label generation or refinement.}

\begin{table}[]
\caption{Online A/B test results. ``WT'' denotes the metric Watch Time, and ``AU'' denotes the metric App Usage.}
\vspace{-10pt}
\label{tab::abtest}
\resizebox{0.4\textwidth}{!}{%
\begin{tabular}{lcccc}
% \hline
\toprule
\multicolumn{1}{c}{\multirow{2}{*}{Business Scenarios}} & \multicolumn{2}{c}{Featured-Video Tab} & \multicolumn{2}{c}{Slide Tab} \\ \cline{2-5}
\multicolumn{1}{c}{}                                    & WT         & AU         & WT     & AU    \\ \hline
WLR                                                   & -                  & -                 & -              & -            \\
D2Q                                                   & +0.273\%           & +0.114\%          & +0.358\%       & +0.135\%     \\
DML($y_{wpr}$)                                          & +0.694\%           & +0.207\%          & +0.648\%       & +0.228\%     \\
DML($y_{wpr}^d$)   & +1.048\%           & +0.332\%          & +1.080\%       & +0.412\%     \\
DML($y_{wpr}^d$,$\{y_{ev}^d, y_{ev}^v, y_{ev}^u\}$)   & +2.230\%           & +0.773\%          & +2.057\%       & +0.709\%     \\
DML($y_{wpr}^d$,$\{y_{lv}^d, y_{lv}^v, y_{lv}^u\}$)   & +1.549\%           & +0.566\%          & +1.284\%       & +0.555\%     \\ 
\bottomrule
\end{tabular}
}
\vspace{-15pt}
\end{table}

\vspace{-5pt}

\section{Conclusion}
\bym{This study delves into leveraging watch time for improved and unbiased modeling of user preferences. We introduce an innovative causal labeling framework, DML, which employs watch time to generate various labels encompassing diverse semantic aspects while incorporating debiasing techniques. Both online and offline experiments are executed on real-world data, offering valuable insights into our approach's efficacy. Results highlight DML's potential in enhancing online video consumption, with its deployment already underway in Kuaishou. However, the current label definition and debiasing process rely on prior knowledge and human expertise. In the future, we plan to develop automated methods to streamline this and align better with online recommendations.}
\begin{acks}
    This work is supported by the National Natural Science Foundation of China (62272437), and the CCCD Key Lab of Ministry of Culture and Tourism.
\end{acks}

\bibliographystyle{ACM-Reference-Format}
\bibliography{6_reference}

\end{document}